\documentclass[prl,showpacs,letterpaper,twocolumn,groupedaddress,footinbib,amssymb]{revtex4}
\usepackage{latexsym}
\usepackage{amsbsy}
\usepackage{amsmath}
\usepackage{graphicx}
\usepackage{subfig}
\usepackage{caption}

\usepackage{color}

\usepackage[ps2pdf, colorlinks, linkcolor = blue, citecolor = black, filecolor = black, urlcolor = blue]{hyperref}

\begin{document}
\date{\today}

\title{Pattern formation in skyrmionic materials with anisotropic environments}

\author{J. Hagemeister$^{1,}$\hyperref[correspondingAuthor]{*}, E.Y. Vedmedenko$^1$, R. Wiesendanger$^1$}

\affiliation{$^1$Department of Physics, University of Hamburg, D-20355 Hamburg, Germany
}

\begin{abstract}

Magnetic skyrmions have attracted broad attention during recent years because they are regarded as promising candidates as bits of information in novel data storage devices. A broad range of theoretical and experimental investigations have been conducted with the consideration of rotational symmetric skyrmions in isotropic environments. However, one naturally observes a huge variety of anisotropic behavior in many experimentally relevant materials. In the present work, we investigate the influence of anisotropic environments onto the formation and behavior of the non-collinear spin states of skyrmionic materials by means of Monte-Carlo calculations. We find skyrmionic textures which are far from having a rotational symmetric shape. Furthermore, we show the possibility to employ periodic modulations of the environment to create skyrmionic tracks.

\end{abstract}

\maketitle

Magnetic skyrmions were originally proposed and investigated in theoretical studies\cite{Bogdanov1994,Bogdanov1994a} and only recently, they were found experimentally in the bulk and thin films of non-centrosymmetric materials\cite{Pappas2009,Muhlbauer2009,Yu2010,Yu12011,Tonomura2012,Seki2012,Moskvin2013} as well as in ultra-thin magnetic films at crystal surfaces\cite{Heinze2011,Romming2013,Romming2015}. They are particle-like objects with a non-vanishing topological charge that is considered to protect them from continuous transformation into the saturated magnetic state. Together with their small lateral size in the nanometer range, this makes skyrmions an interesting candidate for bits of information and information carriers in novel data storage devices \cite{Kiselev2011,Fert2013}.
Typically, magnetic skyrmions form due to the competition of the Dzyaloshinskii-Moriya (DM) interaction\cite{Dzyaloshinskii1958,Moriya1960} and the exchange interaction in the presence of an external magnetic field. The DM interaction can provide a non-vanishing contribution in the presence of inversion asymmetry combined with a large spin-orbit interaction. The properties of skyrmions have both theoretically and experimentally mainly been investigated in the context of materials providing an isotropic environment up to now which leads to the formation of skyrmions with a rotational symmetric equilibrium shape. There are only few studies that deal with the influence of anisotropic environments and only recently, the experimental observation of skyrmions which are deformed with respect to the rotational symmetric shape has been reported for chiral magnets with crystal lattice strain\cite{Shibata2015}.

 However, various experimentally feasible materials naturally exhibit anisotropic behavior for multiple reasons. In the discussion of the origins, we focus on the material systems that allow for interface induced skyrmionic states, only. These material systems typically consist of a single or multiple  atomic, magnetic layers of different atomic species which are deposited succeedingly onto a non-magnetic supporting crystal.
The combination of materials with different lattice constants can give rise to lattice strain and reconstructions in the magnetic surface layers which then exhibit anisotropic environments as has been discussed only recently for the double\cite{Hsu2016a} and triple\cite{Hsu2016} atomic layers of Fe on Ir(111). 
Therefore, the influence of anisotropic environments onto the behavior of skyrmionic states is already interesting for practical reasons. More intriguing is, especially with respect to technological applications, to study the capability of tailoring the properties of skyrmions such as their shape and lateral position by anisotropic environments.  

Here, we report on Monte-Carlo investigations of the magnetic pattern formation in systems with modulated environments and show skyrmionic states which exhibit shapes very different from the rotational symmetric ones that have been discussed in previous studies. Also, we discuss the possibility to align skyrmions along tracks and relate it to recent investigations conducted on the double and triple layers of Fe on Ir(111)\cite{Hsu2016a,Hsu2016}.

We describe a skyrmionic ultra-thin magnetic film by the standard effective Hamiltonian

\begin{multline}
H = - \sum_{<i,j>}J_{ij}\textbf{S}_{i} \cdot \textbf{S}_{j} - \sum_{<i,j>} \textbf{\textit{D}}_{ij} \cdot \left( \textbf{S}_{i} \times \textbf{S}_{j} \right) \\ 
+ \sum_{i}{K_i(\textbf{e}_{K_i}\cdot \textbf{S}_i)^2}- \mu\sum_{i}{\boldsymbol{B} \cdot \textbf{S}_{i}}
\label{eqn:hamilton}
\end{multline}

consisting of the exchange energy, the DM energy, the magnetocrystalline anisotropy and the Zeeman energy. The classical three-dimensional spins $\textbf{S}_i$ of unit length can rotate freely on the unit sphere. We consider an isotropic nearest neighbor DM interaction strength $|\textbf{\textit{D}}| = D$ and for symmetry reasons we assume DM-vectors which both lie in the plane of the magnetic film and are perpendicular to the connection line between neighboring spins\cite{Crepieux}. The values of $J_{ij}$, $K_i$, $\textbf{e}_{K_i}$, $\mu$ and $\textbf{\textit{B}}$ provide the exchange energy parameter, the anisotropy energy parameter, the direction of the uniaxial anisotropy, the magnetic moment and an external magnetic field.
A detailed phase space of a magnetic film which can be described by an Hamiltonian of the type given in eq. \ref{eqn:hamilton} with an isotropic exchange interaction can be found elsewhere\cite{Yu2010,Banerjee2014,Keesman2015}. In order to summarize, a spin-spiral state with a fixed rotational sense is the ground state at low temperatures and zero external magnetic field. The application of a perpendicular external magnetic field can cause the system to exhibit a skyrmionic state at an intermediate field strength or the saturated ferromagnetic state at a sufficiently large field. 

We use two-dimensional triangular lattices with the lattice constant $a$ and mimic the behavior of multiple magnetic layers by employing effective nearest-neighbor exchange interaction parameters $J_{ij}$ similar to previous investigations of the orientation of domain walls \cite{Vedmedenko2004,Vedmedenko2005}. 
However, we go beyond the earlier model and modulate the strength of the exchange coupling not only as a function of the orientation of the respective bond but as a generalization also of the position in the lattice according to the following description

\begin{align}
& J_{ij}^2(\textbf{r}_{i},\textbf{r}_{j}) = \left( j_\mathrm{M}^2 \cdot \frac{|\textbf{e}\cdot \textbf{r}_{ij}|^2}{a^2} + j_\mathrm{m}^2 \cdot \frac{|\textbf{e} \times \textbf{r}_{ij}|^2}{a^2}  \right) \label{eqn:J}\\
& \textbf{e}^T (\alpha) = (\cos(\alpha), \sin(\alpha),0) \\
& \alpha  = \alpha_\mathrm{max}\cdot \sin\left( \frac{2 \pi}{\lambda} \cdot \left[\frac{1}{2}(\textbf{r}_{i} + \textbf{r}_{j})\cdot \textbf{e}_{[10\overline{1}]}\right] \right)  
\label{eqn:modulation}
\end{align}

The equation \ref{eqn:J} is equivalent to the parametrization of an ellipse with the semi-major axis $j_\mathrm{M}$ and semi-minor axis  $j_\mathrm{m}$ in the case of $j_\mathrm{M} > j_\mathrm{m}$. The ellipse lies within the plane of the magnetic film and the direction of the major axis $\textbf{e}$ is locally rotated by the angle $\alpha$ with respect to the $[10\overline{1}]$ direction. $\textbf{\textit{r}}_{ij}$ is the vector connecting the lattice sites $i$ and $j$. The choice of unequal pa\-ra\-me\-ters $j_\mathrm{M}$ and $j_\mathrm{m}$ provides an anisotropic interaction strength which then depends on the crystallographic direction. The angle $\alpha_\mathrm{max}$ and the period $\lambda$ in units of $a$ can be used to create a periodic modulation of the anisotropic exchange environment along the crystallographic direction $[10\overline{1}]$.

\begin{figure}[ht!]
\centering
\includegraphics[width=8.2cm]{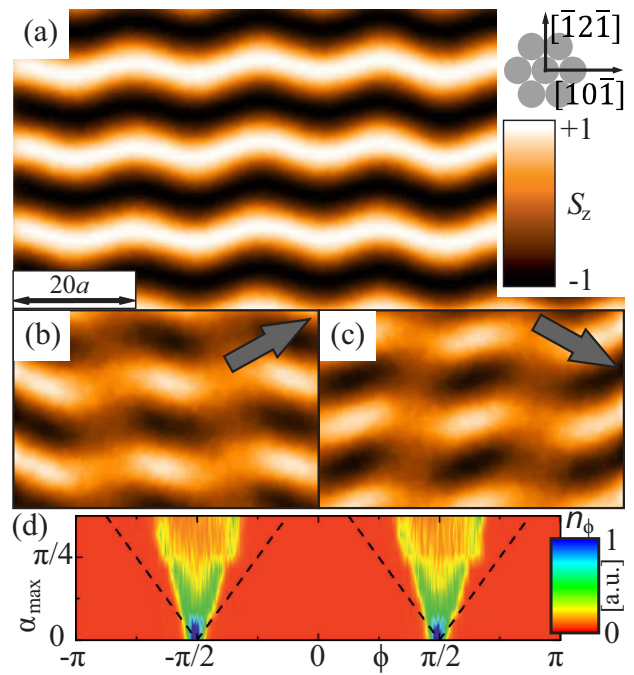}
		\caption{Perpendicular (a) and in-plane magnetic contrasts (b,c) of a spin-spiral state with wiggling wave fronts on a triangular lattice of $9700$ sites with periodic boundary conditions in the $[\overline{1}2\overline{1}]$ direction. The arrows in (b),(c) give the direction for the in-plane contrasts. (d) The frequency of the azimuthal angle $\phi$ with respect to the $[10\overline{1}]$ direction.}
		\label{graphic:fig1}
\end{figure}

Fig.~\ref{graphic:fig1}a shows the perpendicular component of the magnetization of a spin-spiral state for the modulation period $\lambda = 20\,a$ and modulation angle $\alpha_\mathrm{max} = 29^{\circ}$. The interaction strengths are $j_\mathrm{m}/j_\mathrm{M} = 0.5$ and $j_\mathrm{m}/D = 1.6$ and the anisotropy is $K=0$. The temperature of the system was decreased from $k_\mathrm{B}T/D = 1.7$ to $k_\mathrm{B}T/D = 8.6\cdot 10^{-3}$ using $20$ temperature steps with $10^5$ Monte-Carlo steps each ensuring that the global energy minimum is reached. We used a single spin update mechanism based on the Metropolis algorithm. The resulting spin-spirals have a wave length of approximately $16.5\,a$. The wave fronts are on average aligned in parallel to the $[10\overline{1}]$ direction and the periodicity of their zigzag pattern is in agreement with the modulation period of the exchange energy.
Part of the spin contrast vanishes (Fig.~\ref{graphic:fig1} b,c) when calculating the projection of the in-plane component of the magnetization onto the directions which enclose an angle of $\pm \alpha_\mathrm{max} = \pm 29^{\circ}$ with the $[10\overline{1}]$ direction. This finding can be verified by the investigation of the incidence $n_{\phi}$ of the spins' azimuthal angles $\phi$ with respect to the $[10\overline{1}]$ direction which is shown in Fig.~\ref{graphic:fig1} d. With a vanishing modulation amplitude $\alpha_\mathrm{max} = 0$, the spins are aligned in the plane given by the system's normal vector and the $[\overline{1}2\overline{1}]$ direction. An increase of the modulation amplitude $\alpha_\mathrm{max} $ causes the single peaks at $\pm 0.5\,\pi$ to split into double peaks with a distance of approximately $\alpha_\mathrm{max}$. The present model seems to capture the main features of the spin-spiral states which have been observed experimentally in the double and triple atomic layers of Fe on Ir(111)\cite{Hsu2016a,Hsu2016}.    
 
\begin{figure}[ht!]
\centering
\includegraphics[width=8.2cm]{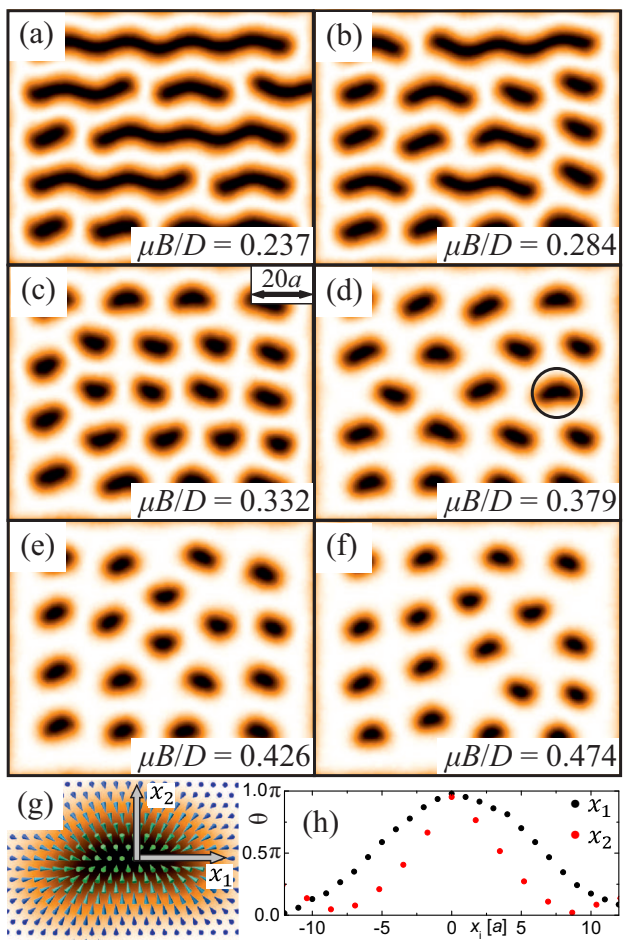}
		\caption{(a-f) Perpendicular contrast of the magnetic pattern which is obtained under application of a perpendicular magnetic field to a zigzag spin-spiral state with open boundary conditions. (g) Spin texture of the magnetic object marked in (d). (h) Profiles of the polar angle for the magnetic texture in (g).}
		\label{graphic:fig2}
\end{figure}

However, it is more intriguing to explore the influence of the modulation of the exchange energy onto the skyrmion state. Fig.~ \ref{graphic:fig2} (a-f) show the magnetic ground states which are obtained when the temperature of the system is reduced with applied constant magnetic fields. Except for the magnetic field, all simulation parameters are chosen to be the same as before. The spin-spirals break up at the magnetic field $\mu B/D = 0.237$ and elongated zigzag structures are created which eventually shrink at magnetic field strengths larger than $\mu B/D = 0.284$. No individual magnetic objects are observed for fields larger than $\mu B/D = 0.62$. The magnetic objects possess a unique rotational sense and yield a non-vanishing topological charge $Q$ which is usually defined via

\begin{displaymath}
Q = \frac{1}{4 \pi}\int_{A}{\textbf{m}\cdot \left( \frac{\partial \textbf{m}}{\partial x} \times \frac{\partial \textbf{m}}{\partial y} \right) \mathrm{d}x \mathrm{d} y }
\end{displaymath}

as an integral over the magnetic surface $A$ whose local direction of magnetization is described by the continuous field $\textbf{m}$. For the present investigations, an adoption of the formula for discrete systems is used. 
Furthermore, the magnetic objects are non-rotational symmetric as investigated exemplarily for the marked spin texture in Fig.~\ref{graphic:fig2}d. Fig.~\ref{graphic:fig2}g,h show the atomic spin configuration and the corresponding line profiles of the spins' polar angles. The profiles of the polar angle of the local magnetization direction clearly show an elongation of the structure into the $[10\overline{1}]$ direction compared to the $[\overline{1}2\overline{1}]$ direction.

\begin{figure}[ht!]
\centering
\includegraphics[width=8.2cm]{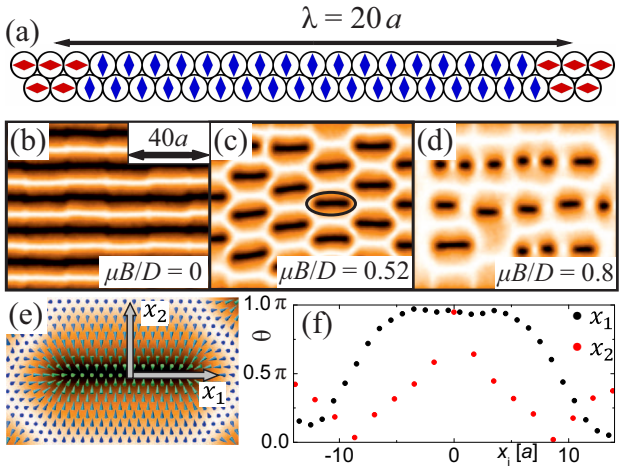}
		\caption{(a) Sketch of the atomic lattice in the top view indicating regions of easy in-plane anisotropy axes parallel to the $[10\overline{1}]$ (red) and $[\overline{1}2\overline{1}]$ (blue) direction. (b)-(d) Perpendicular magnetic contrast of the ground state in an applied magnetic field. (e) Spin configuration of a skyrmion marked in (c). (f) Profiles of the polar angle through the magnetic texture shown in (e).}
		\label{graphic:fig3}
\end{figure}

So far, we have focused on the effects of a spatial modulation of the exchange energy parameter in skyrmionic materials and neglected the anisotropy energy.
However, from previous studies it is known that the anisotropy energy can in principle vary locally in thin magnetic films\cite{Porrati2004}. In the following, we investigate the effects of a spatial modulation of the anisotropy energy in combination with an isotropic exchange and DM interaction. We introduce stripe-like regions with an easy in-plane axis parallel to the $[\overline{1}2\overline{1}]$ direction and easy in-plane axis parallel to the $[10\overline{1}]$ direction of the widths $dl_1 = 18\,a$ and $dl_2 = 2\,a$ which are periodically repeated along the $[10\overline{1}]$ direction with a periodicity of $20\,a$. In Fig.~\ref{graphic:fig3} a, a sketch of the atomic lattice is given indicating the anisotropy axis for each lattice site. 
The same anisotropy energy constant of $K/D=0.6$ is assumed for the two regions and $J/D = 2.1$ is chosen. 
Fig~\ref{graphic:fig3} b)-d) show the magnetic ground states for different magnetic fields after the reduction of the temperature in the same fashion as before. The spin-spiral period at zero magnetic field is approximately $16\,a$ and the wave fronts are aligned in parallel to the $[10\overline{1}]$ direction.
With an applied magnetic field of $\mu B/D = 0.52$, an ordered state of elongated magnetic objects is formed. Their centers lie in the middle of the regions that possess an anisotropy axis parallel to the $[\overline{1}2\overline{1}]$ direction and their elongation into the $[10\overline{1}]$ direction is determined by the areas in which the anisotropy axis is parallel to the $[10\overline{1}]$ direction. This finding is verified exemplarily for a single magnetic object which is highlighted in Fig.~\ref{graphic:fig3}c. Fig.~\ref{graphic:fig3} e shows the corresponding spin configuration and Fig.~\ref{graphic:fig3} f displays the profiles of the polar angle of the magnetization along two perpendicular crystallographic axes. One can clearly discern a lateral size of the magnetic object parallel to the $[10\overline{1}]$ direction of about $20\,a$. In other words, the modulation of the anisotropy energy aligns the magnetic objects on tracks along the $[\overline{1}2\overline{1}]$ direction. 
 For magnetic fields larger than $\mu B/D \approx 0.7$, the magnetic objects split into two smaller parts which themselves vanish for fields larger than  $\mu B/D \approx 0.9$.

For technological applications, it would be desirable to align skyrmions on tracks along which they then may be moved for example via a spin-polarized current\cite{Fert2013,Iwasaki2013,Zhou2014,Purnama2015}. This may be realized by producing long narrow stripes of magnetic materials along which the skyrmions can propagate. Another idea would be to use anisotropic environments due to lattice strain in pseudomorphically grown ultrathin magnetic films. An experimentally relevant system may be the recently investigated triple layer of Fe on Ir(111)\cite{Hsu2016}.

\begin{figure}[ht!]
\centering
\includegraphics[width=8.2cm]{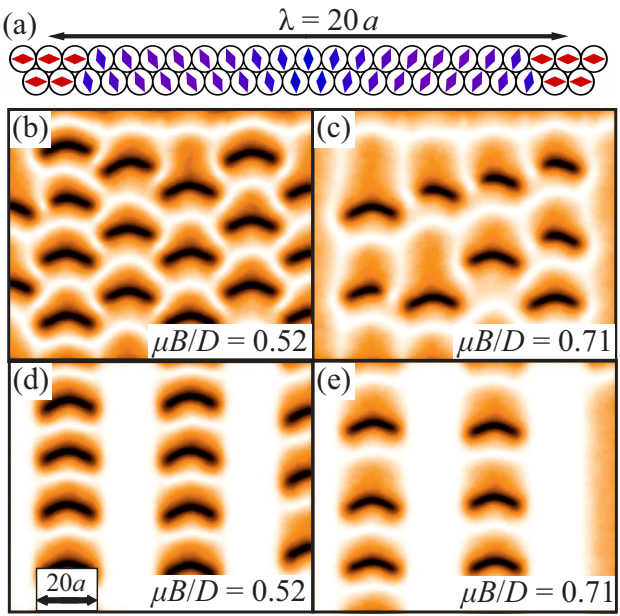}
		\caption{(a) Sketch of the atomic lattice in the top view indicating the spatial modulation of the direction of the easy anisotropy axis. (b),(c) Perpendicular contrast of the magnetic ground states at different magnetic fields. The spatially modulated anisotropy energy is combined with the anisotropic exchange interaction used in Fig.~\ref{graphic:fig1},\ref{graphic:fig2} and an isotropic DM-interaction. (d),(e) Formation of skyrmion tracks by a spatial modulation of local magnetic anisotropy as explained in the text.}
		\label{graphic:fig4}
\end{figure}

In the following, we want to gain insight into the formation of linearly aligned skyrmionic structures along a track by a spatial modulation of the energy parameters in a 2D model system. For this purpose, we combine the effects of the modulation of the exchange and anisotropy energy with an isotropic DM-interaction on the magnetic ground state of skyrmionic systems. The parameters for the exchange and DM-interaction are chosen as in the simulations for Fig.~\ref{graphic:fig1}, \ref{graphic:fig2}. Additionally, we assume a spatial modulation of an easy in-plane anisotropy axis with the energy $K/D = 0.6$ similar to simulations for Fig.~\ref{graphic:fig3} but partly adopt the anisotropy axes to follow the modulation of the exchange interaction as indicated in Fig.~\ref{graphic:fig4}a. For two magnetic fields, the magnetic ground states at low temperatures are shown in Fig.~\ref{graphic:fig4}b,c. One observes ordered bent non-collinear spin states with a non-vanishing topological charge at the magnetic field $\mu B/D = 0.52$ which vanish at magnetic fields larger than $\mu B/D \approx 0.71$. As before, the magnetic objects are aligned along tracks parallel to the $[\overline{1}2\overline{1}]$ direction. However, the magnetic objects of neighboring tracks influence each other in the current status in such a way that they could not be moved independently from each other parallel to the $[\overline{1}2\overline{1}]$ direction. We exclude every second track by locally substituting the easy in-plane axis by an easy out-of-plane anisotropy axis with the energy $K/D = 0.6$. The resulting magnetic equilibrium skyrmion-like patterns are confined to spatially separated tracks parallel to the $[\overline{1}2\overline{1}]$ direction (Fig.~\ref{graphic:fig4}d,e).

To summarize, we have shown possible influences of spatial modulations of the exchange interaction and anisotropy energy onto the formation of non-collinear spin-states in skyrmionic materials. Initially, magnetic skyrmions were introduced as rotational symmetric objects with a non-vanishing topological charge that form in the presence of DM interaction, the exchange interaction and the magnetocrystalline anisotropy. In contrast to bubble domains, they possess an equilibrium size. Since the magnetic objects that we have obtained possess the topological charge $Q=1$ and have equilibrium sizes $\lesssim 10\,\mathrm{nm}$, they exhibit indeed the main characteristics of skyrmions.

\section*{Acknowledgments}
We thank  Pin-Jui Hsu, Niklas Romming, Lorenz Schmidt, Aurore Finco, Andr\'e Kubetzka and Kirsten von Bergmann for fruitful discussions.
Financial support from the DFG in the framework of the SFB668 as well as from the European Union (FET-Open MagicSky No. 665095) is gratefully acknowledged.

\vspace{1cm}

{*Corresponding author.\label{correspondingAuthor}}
 
jhagemei@physnet.uni-hamburg.de


\end{document}